\documentclass[a4paper,fleqn,usenatbib]{mnras}
\usepackage{color}
\usepackage{graphicx}	
\usepackage{amsmath}	
\usepackage{amssymb}	

\newcommand\simlt{\lower.5ex\hbox{$\; \buildrel < \over \sim \;$}}
\newcommand\simgt{\lower.5ex\hbox{$\; \buildrel > \over \sim \;$}}

\title[Limits on the Growth Rate of SMBHs]{Limits on the growth rate of supermassive black holes at early cosmic epochs}
\author[A. Levinson \& E. Nakar]{Amir Levinson \& Ehud Nakar
\\
The Raymond and Beverly Sackler School of Physics and Astronomy, Tel Aviv University, Tel Aviv 69978, Israel\\
}

\begin{document}
\maketitle
	
\begin{abstract}
The effect of AGN wind feedback on the accretion rate and mass evolution of supermassive black  holes (SMBH) is considered. It is shown, under reasonable assumptions, that the rate at which gas can be supplied to a SMBH at the center of a young galaxy, is limited to
$\sim 20 (\sigma/350~ {\rm km ~s^{-1}})^4~M_\odot/yr$ (where $\sigma$ is the velocity dispersion of the host bulge)
by interaction of winds expelled from the innermost regions
of the accretion flow with the gas in the bulge.  This rate is independent of the black hole mass but is sensitive to the properties of the host bulge. It is further argued that the interaction of the wind and the inflowing gas in the
bulge can strongly affect the structure of the accretion flow in the super-Eddington regime, potentially leading
to highly super-Eddington accretion into the SMBH. About 300 Myr after the onset of the accretion phase, the AGN wind expels all the gas from the bulge and the accretion rate is strongly suppressed.
This scenario is in remarkable agreement with recent observations, as it (i) reveals the origin of the maximal observed accretion rates, (ii) accounts for the inferred growth rate of SMBHs at high
redshift (independent of their initial seeds masses), (iii) elucidates the decline in the accretion rate at lower redshifts, and (iv) explains the relation between $\sigma$  and the mass $M_{BH}$ of the central SMBH, measured in the local Universe.
\end{abstract} 


\section{Introduction}

\begin{figure}
\centering
\includegraphics[width=1.1\columnwidth]{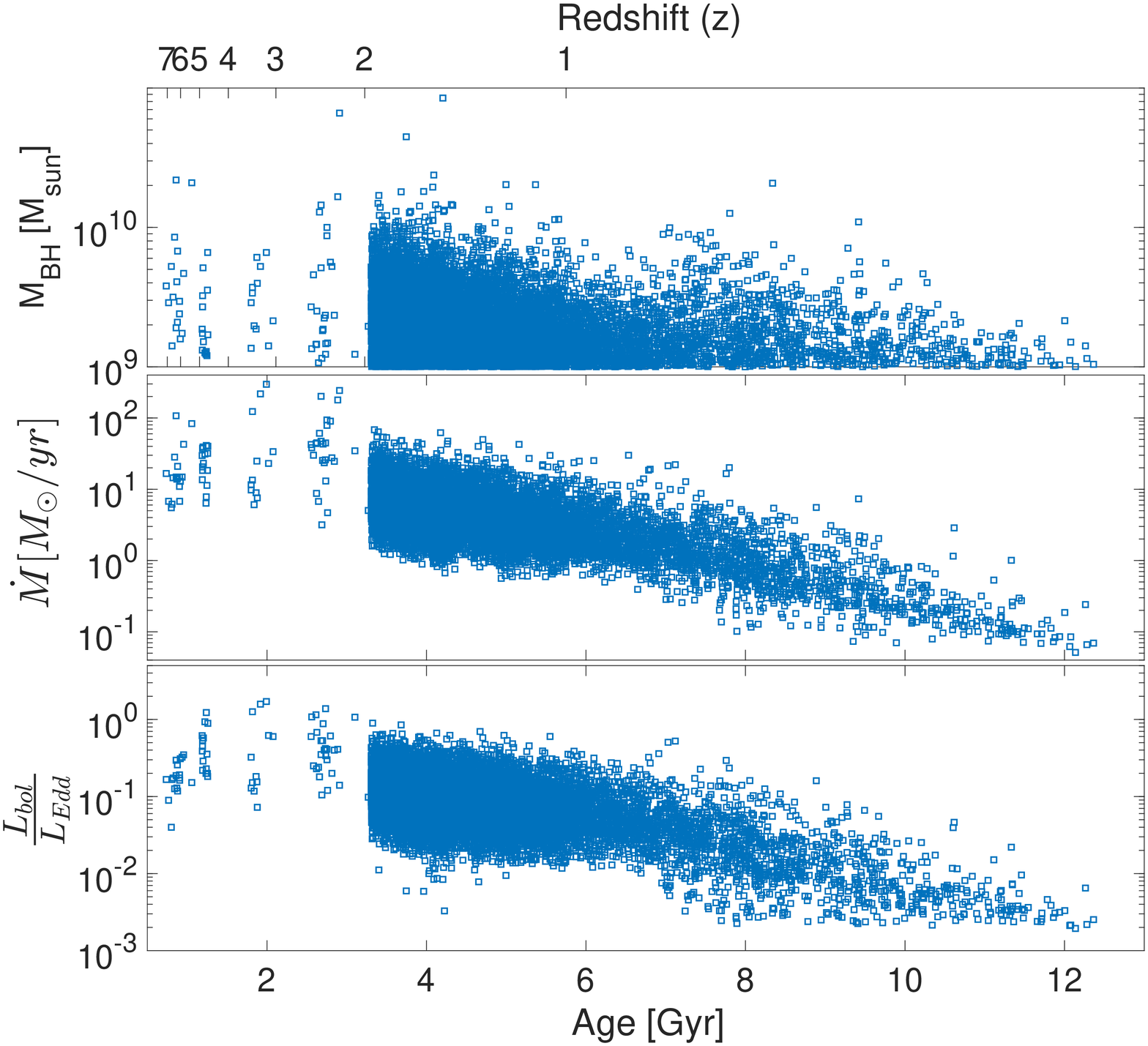}
\caption{Masses (top), accretion rates (middle) and Eddington ratios (bottom) of observed SMBHs with masses larger than
$10^9~M_\odot$, as a function of cosmic time. The observational data is compiled from the litrature
\citep{trakhtenbrot2011,trakhtenbrot2012,trakhtenbrot2017}. 
The accretion rates are calculated based on the bolometric luminosity, 
assuming a radiative efficiency of 0.1. The upper panel indicates that the most massive SMBHs obtained masses of $10^{9}-10^{10}~M_\odot$ when the Universe was only $\sim 800$ Myr old, 
and the middle panel indicates that these SMBHs accrete at a rate of a few tens of $M_\odot/yr$. 
This trend continues up to a cosmic time of $\sim 3$ Gyr (z=2), after which the accretion rate starts to drop significantly.
}   
\label{fig:f1}
\end{figure}

The growth of supermassive black holes (SMBHs) in the young Universe, 
and the effect of the associated quasar activity on the evolution of the host galaxy,
are central issues in the theory of galaxy formation and evolution. 
Black holes with masses  $10^{9}-10^{10}~M_\odot$, accreting at a roughly constant rate of
a few tens of $M_\odot/yr$, have been detected at redshifts from $z\simeq2$ up to $z\simeq 7$,  
when the Universe was less than $1$ Gyr old \citep{kurk2007,willott2010,trakhtenbrot2017}.  
Below $z\simeq2$, the  accretion rate appears to decrease with increasing 
cosmic time \citep{trakhtenbrot2011,trakhtenbrot2012}.   These trends are clearly seen in figure \ref{fig:f1} 
that exhibits compilation of data from the literature for AGNs at redshifts $0 \le z \le 7$, harbouring SMBHs of masses larger than $10^9 M_\odot$.  What determines the observed accretion rate?

Two different limits on the accretion rate are commonly discussed in the literture. 
The first one is the rate at which material with sufficiently low angular momentum 
can be transferred from galactic scales into the vicinity of the SMBH.  
This rate is presumably  determined by conditions far from the black hole 
and cannot exceed the spherical Bondi-Hoyle-Lyttleton accretion rate \citep{hoyle1939,bondi1952}:  

\begin{equation}
\dot{M}_{max}=4\pi \rho_g r^2 \sigma=2f_g\sigma^3/G \approx 4\times10^3\, \sigma^3_{350}\quad M_\odot\, yr^{-1},
\label{dotMmax}
\end{equation}
where following \cite{silk1998}, we model a protogalaxy as an isothermal sphere of dark matter, stars and gas
with a constant velocity dispersion $\sigma = 350\, \sigma_{350}$ km s$^{-1}$, a gas fraction $f_g\sim0.2$ and a gas density $\rho_g=f_g \sigma^2/2\pi G r^2$, here $G$ is the gravitational constant and $r$ is the distance from the SMBH.
The second accretion rate limit that is often considered is the Eddington limt,
\begin{equation}
\dot{M}_{Edd} =  70 M_{BH,9.5}\quad M_\odot\, yr^{-1} ,
\label{dotMEdd}
\end{equation}
above which the radiation emitted by the accretion flow becomes trapped, potentially blowing away a significant fraction 
of the inflowing gas before it gets absorbed by the black hole.  This limit
depends only on the BH mass, $M_{BH}=10^{9.5} M_{BH,9.5}~M_\odot$, and is independent of the properties of the host galaxy.  
It has been pointed out that it may be surpassed under certain conditions \citep{volonteri2015,begelman2017}.

None of these limits is consistent with the observed accretion rates depicted in figure \ref{fig:f1}. The Bondi limit is much higher than the  observed rates, while if the accretion rate of SMBHs at high redshifts would have been dictated by the Eddington limit we would expect the Eddington ratios of sources at $z>2$ in the bottom panel of figure \ref{fig:f1} to be concentrated around unity.   In this paper we propose that throughout its growth phase, the rate at which a SMBH is accreting matter is limited by the action of winds, expelled from the inner regions of the accretion disk surrounding it, on inflowing gas at much larger scales.  This self-regulation of black hole accretion has been largely ignored until now, albeit it has been recently found to be the dominant limiter in numerical simulations on galactic scales that resolve the SMBH sphere of influence \citep[e.g.,][]{negri2017}.

In \S\ref{sec:mdot} we derive analytically (under simplifying assumptions) an upper limit on the accretion rate, by accounting for the feedback of winds ejected during the accretion process.  We show that under reasonable conditions this limit depends primarily on the bulge properties, and is in agreement with the observed accretion rates exhibited in figure \ref{fig:f1}.

In \S\ref{sec:supercrit} we argue that previous considerations of Eddington limited accretion ignored the interaction between the wind and the gas in the bulge. It is commonly argued \citep[e.g.,][]{shakura1973,begelman2012} that supercritical accretion flows  eject  most of their mass at large radii, before it reaches the black hole, in which case the SMBH growth rate might be suppressed. We show that the wind-bulge interaction may apply a significant pressure on the disk. Such pressure, if applied, prevents ejection of mass from the disk far from the SMBH and enforce highly super-Eddington accretion, thereby alleviating the "seed mass problem" \cite[][ and references therein]{volonteri2010}.

In \S\ref{sec:Msigma} 
we determine the final mass of the SMBH, following similar arguments to those of \cite{silk1998} and \cite{king2003}. 
They have shown that  the  $M_{BH}-\sigma$ relation \cite[ and references therein]{kormendy2013} can be reproduced under the assumption 
that either the wind power \citep{silk1998} or wind momentum flux (King 2003) are proportional to the Eddington luminosity of the SMBH at 
any given time.  This has later been confirmed by numerical simulations that incorporated such feedback \citep[e.g.,][]{dimatteo2005}.  
In difference, our analysis implies that the feedback regulated rate is not related to the Eddington limit.  
We show that the time it takes the wind to expel the gas in the bulge and terminate accretion is approximately 300 Myr, provided that the shocked wind cools rapidly.  The final black hole mass thereby obtained is consistent with the observed $M_{BH}-\sigma$ relation, albeit different than the relations previously derived.

\section{Regulation of accretion rate by mechanical feedback}\label{sec:mdot}
As will be shown below, the regulated accretion rate depends on the wind momentum, which we parametrize 
in terms of the ratio between the outflow mass flux $\dot{M}_w$ and the mass accretion 
rate $\dot{M}_{BH}$, 
\begin{equation}
\alpha=(\dot{M}_{BH}/\dot{M}_w)^{1/2}, 
\end{equation}
and the wind energy extraction efficiency, 
\begin{equation}
\epsilon=\dot{M}_w v_w^2/(2\dot{M}_{BH}c^2),
\end{equation}
where $v_w=c\, \alpha\sqrt{2\epsilon}$ is the wind terminal velocity and $c$ is the speed of light.  
Observations of outflows from AGNs accreting at mildly sub-Eddington rates \citep{borguet2013,cicone2014,chamberlain2015,williams2016} indicate wind efficiencies of about $\epsilon \sim 10^{-3}-10^{-2}$, and wind velocities $v_w \sim 0.1 c$ \citep{borguet2013,chamberlain2015,williams2016,parker2017}, corresponding to $\alpha\sim1$.   In the regime of super-Eddington accretion, where no observational constraints are available, general relativistic radiation magnetohydrodynamical simulations show that a fixed fraction of the accretion energy is extracted
from the accretion flow and emerges in the form of super-Eddington winds  \citep{sadowski2016,sadowski2016b}.   For a non-rotating black hole the simulations 
yield $\epsilon\simeq 0.03$ and $\alpha\sim 1$, independent of accretion rate  and black hole mass.  
For a rapidly rotating black hole $\epsilon$ may be larger by a factor of about 3. 

\begin{figure}
\centering
\centerline{
\includegraphics[width=0.5\columnwidth, height=4cm]{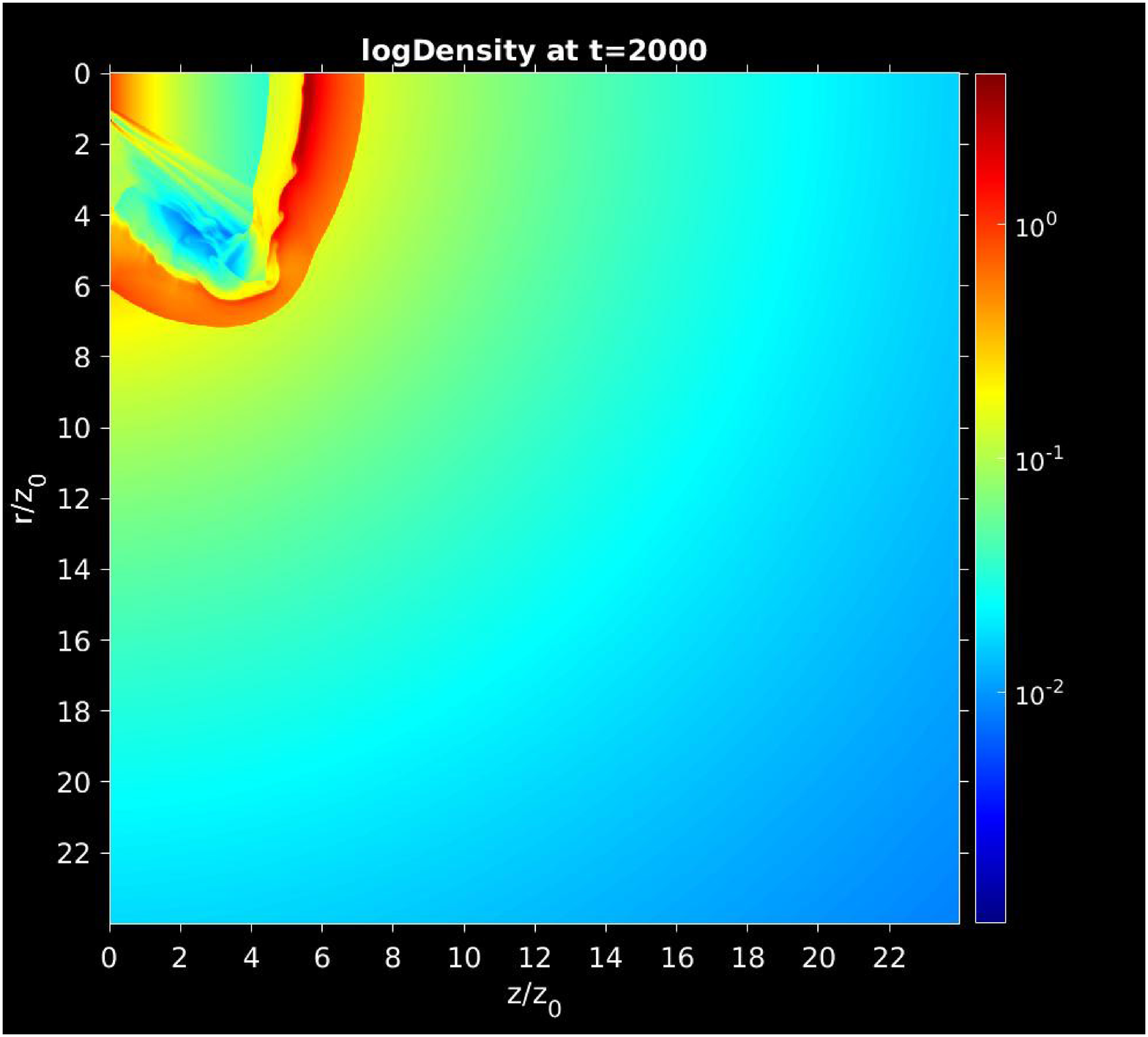}\includegraphics[width=0.5\columnwidth, height=4cm]{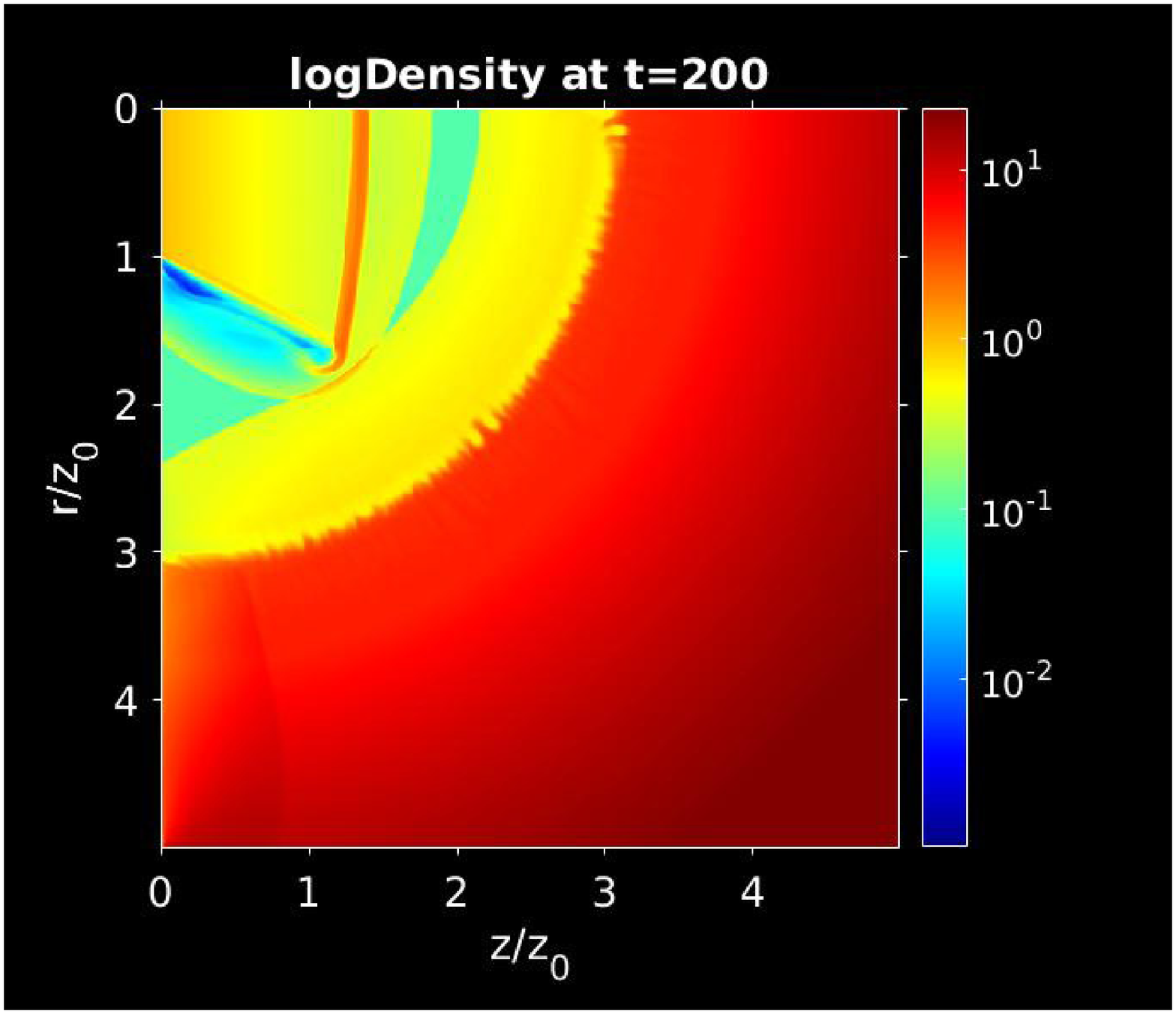}}
\centerline{
\includegraphics[width=0.5\columnwidth, height=4cm]{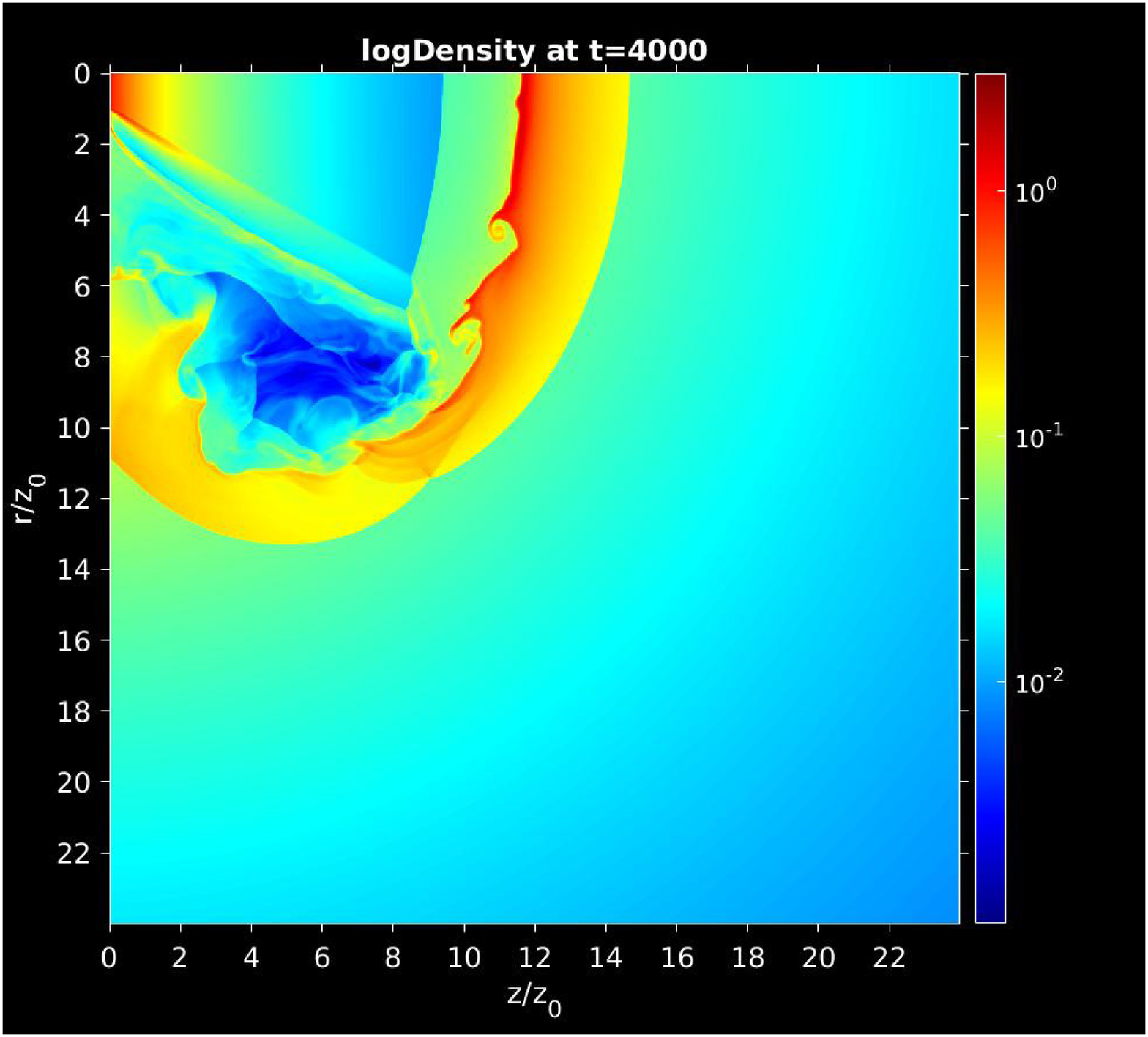}\includegraphics[width=0.5\columnwidth, height=4cm]{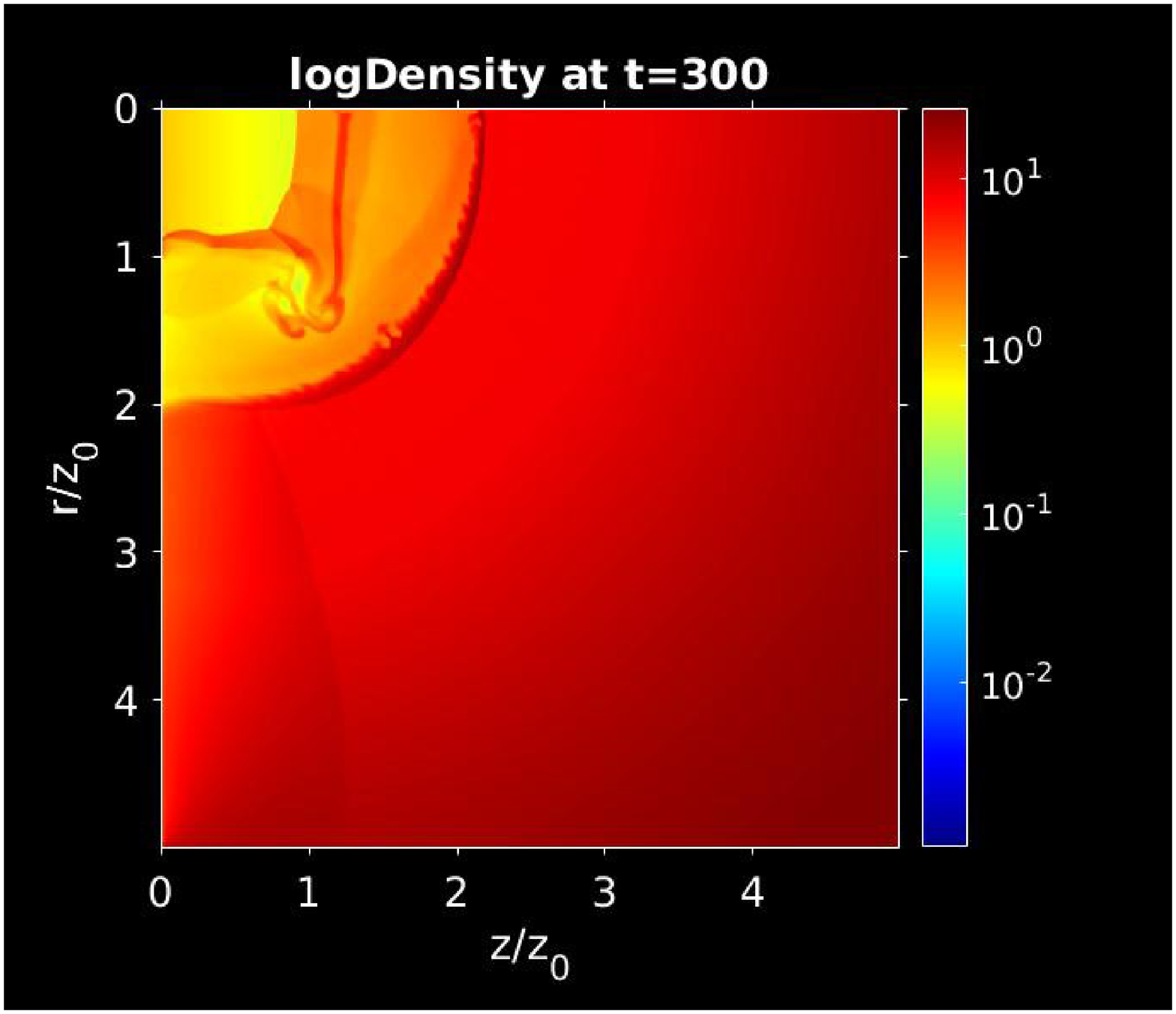}}
\centerline{
\includegraphics[width=0.5\columnwidth, height=4cm]{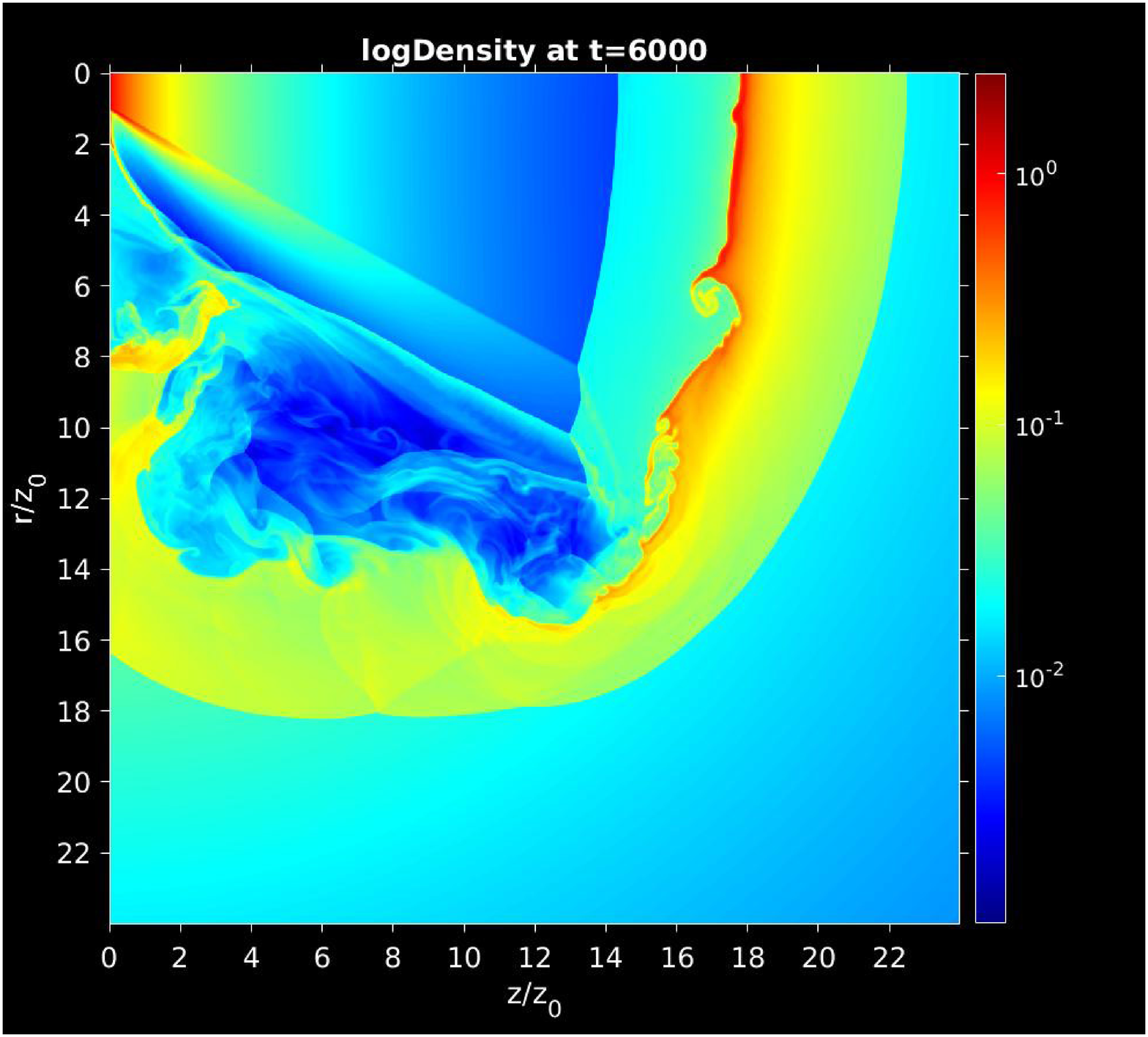}\includegraphics[width=0.5\columnwidth, height=4cm]{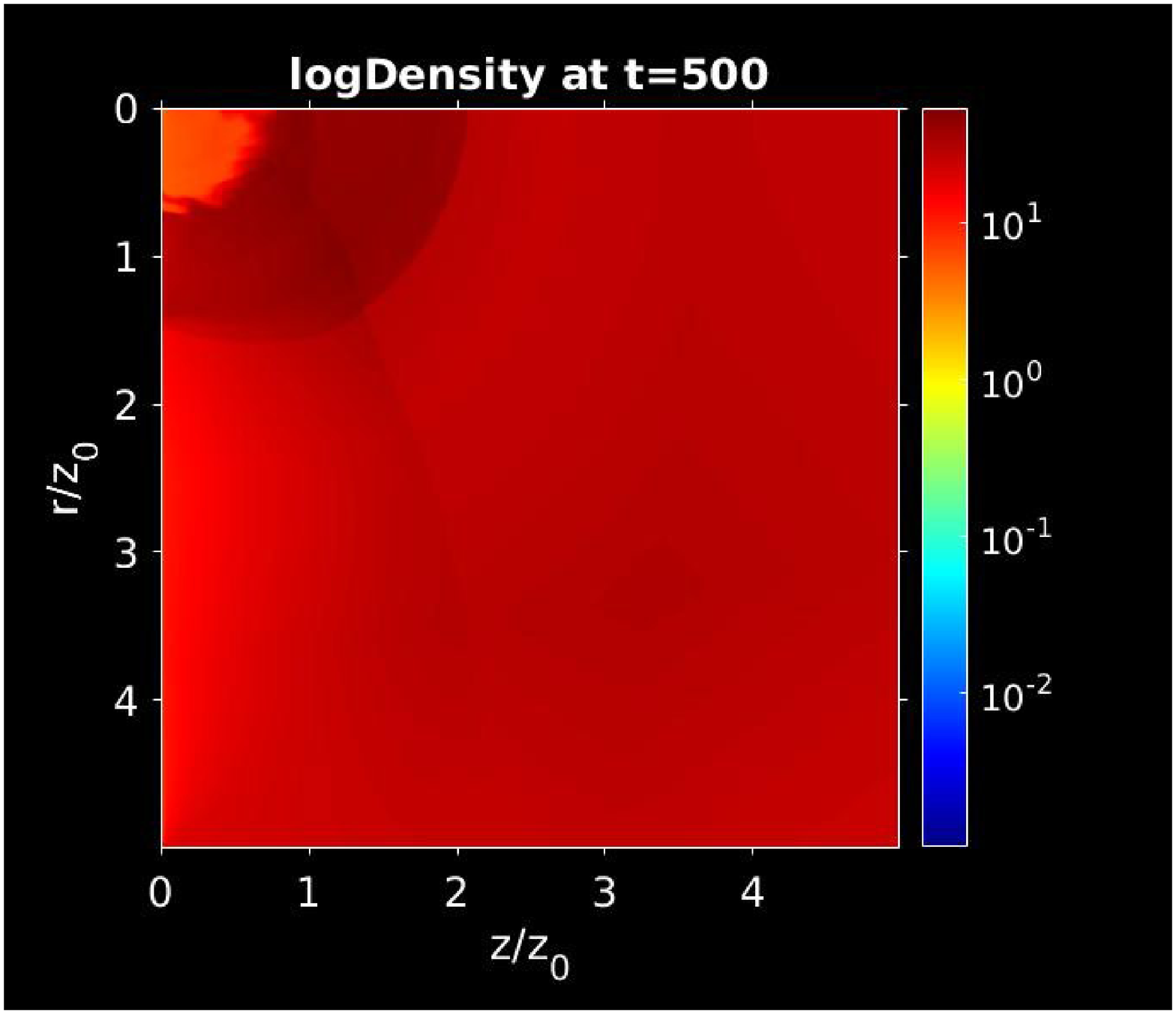}}
\caption{Snapshots from  2D simulations of wind propagation in an external medium.  
The ram pressure of the wind in the simulation shown in the left panels exceeds the pressure of the ambient gas, whereas in the case shown
the right panels the ram pressure of the wind is smaller than the gas pressure.  The opening angle of the injected wind is $45^\circ$ in both cases.
The reddish-yellow stripe seen in the left panels is the expanding shocked bubble (cocoon).  As can be seen, it is nearly spherical.    
For clarity of presentation, we inject the wind in the simulation exhibited in the right panels into a void prior to its collision with the ambient gas.  
The upper right panel shows the moment of collision.  The ratio of gas pressure to wind ram pressure at collision is $p_g/\rho_wv_w^2= 16$.  As seen, 
the wind is suffocated shortly after it collides with the external medium. The simulations were done using the  hydrodynamic code PLUTO \citep{Mignone2007}.}
\label{fig:simulation}%
\end{figure}

If the ram pressure of the wind is lower than the pressure of the ambient gas in the bulge then the wind cannot propagate throug the ambient gas. Instead it inflates a high pressure shocked bubble that ultimately chokes the wind source. The wind in this case does not affect the accretion. If, however, the wind ram pressure is high enough, then it propagates outwards through the inflowing gas, producing an expanding cocoon that contains the shocked wind and shocked ambient medium. The expanding cocoon pushes back the inflowing gas, halting further supply to the black hole, and ultimately quenching the accretion. 
The remaining outflow energy is then exhausted over a few expansion scales, after which the accretion cycle resumes. 
We note that even if the outflow is conical the inflated cocoon is quasi-spherical, as seen in the left panels of        
Figure \ref{fig:simulation}.  This figure presents numerical simulations of the evolution of a wind launched with a half opening  angle of $45^\circ$, in two distinct cases: one where the ram pressure of the wind is larger than the ambient pressure (left panels) and one where it is lower (right panels). The first simulation (left panels) indicates that the quasi-spherical cocoon inflated during a successful wind propagation prevents any infalling material from reaching the BH vicinity.
The second simulation (right panels) shows that the wind quickly suffocates when its ram pressure is lower than 
the pressure of the infalling gas.  

As seen in the left panels of Figure \ref{fig:simulation} the contact becomes 
Kelvin-Helmholtz unstable, owing to shear flows in the shocked wind which are generated by the complex structure of the 
cocoon. In the configuration we simulated the contact is stable to other instability types such as Rayleigh-Taylor, although proper
account of the gravitational filed of the bulge may alter the latter result.
The growth of Kelvin-Helmholtz modes (or other types of modes if unstable) merely leads to mixing of shocked wind and shocked ambient matter, which may affect the cooling (see below), but not the forward shock and the overall dynamics of the cocoon.  Nonetheless, it could well be that some filaments of shocked ambient matter produced by the instability are being pushed in by the gravitational field and ultimately swallowed by the black hole.  This might alter our estimate 
of the regulated accretion limit  in ways yet to be explored.  The results of \cite{negri2017} (see further discussion below) indicate strong suppression 
of the accretion rate by wind feedback, in accord with our heuristic arguments.

The regulated mass inflow rate can be estimated by equating the ram pressure of the wind, $\rho_w v_w^2$, with the ambient gas pressure, $p_g\simeq \rho_g\sigma^2=f_g \sigma^4/(2\pi r^2 G)$ \citep{king2003,Murray2005,king2005}.  This yields, up to a geometrical factor,
\begin{equation}\label{mom_balance}
\frac{f_g\, \sigma^4}{G}= \dot{M}_wv_w \simeq \alpha^{-1}\sqrt{2\epsilon}\dot{M}_{BH} c,
\end{equation}
from which we obtain 
\begin{equation}\label{acrt_rate}
\dot{M}_{BH} \simeq \dot{M}_c\equiv 20 ~\alpha \frac{f_g}{0.2}\left(\frac{\epsilon}{0.01}\right)^{-1/2} \sigma^4_{350}\quad M_\odot \, yr^{-1},
\end{equation}
where typically $\alpha\sim1$ for the type of winds considered here.
Equation \ref{acrt_rate} implies that as long as enough gas can be transferred to the vicinity of a SMBH at the center 
of a young galaxy, it will accrete, on the average, at a
roughly constant rate which is independent of its mass, but is sensitive to the bulge properties (dispersion and gas fraction). If supercritical accretion is possible, which as we argue in the next section may very well be the case, then the accretion rate is only weakly dependent (via $\epsilon$ and $\alpha$) on the accretion mode (sub or supper Eddington).
 
The above analysis ignores the complex structure of the infalling gas, as well as additional processes such as star formation and supernovae feedback.  The main uncertainty that can alter our simplified picture is the multi-phase structure of the gas. 
One might suspect that the dense phase, if exists, will be engulfed by the wind, ultimately reaching the black hole vicinity,
in which case the feedback may be strongly suppressed. 
However, under a wide range of conditions dense clouds will not survive the passage through the shocked bubble driven by the wind-gas interaction.  They will be disrupted by  Kelvin-Helmholtz and Rayleigh-Taylor instabilities roughly over the time it takes the shock to cross the cloud.  For a cloud of a typical size $R_{cloud}$ and density $\rho_{cloud}$, the shock crossing time is roughly $t_{cross}\simeq (\rho_{cloud}/\rho_b)^{1/2}(R_{cloud}/v_b)$, where $\rho_b\simeq \rho_g$ and $v_b$ are the density and velocity of the expanding shocked bubble. The cloud is disrupted and advected with the outflow if $t_{cross}$ is much shorter than the flow time $t_f = r/v_b$, where $r$ is the distance of the cloud from the center, that is, if  $R_{cloud}/r \ll  (\rho_g/\rho_{cloud})^{1/2}$. Given that we expect $R_{cloud} \ll r$, even clouds with a considerable density contrast will be disrupted. 

Star formation is not expected to play a major role, since it does not affect the accretion nor the feedback, unless it consumes all the gas. Given that the highest star formation rates observed at high redshift are $\sim 1000 {\rm~M_\odot/yr}$ \citep[e.g.,][]{barger2014} this cannot happen on the time scale over which the wind bubble expands ($\ll 1$ Gyr). The same star formation rate implies also that the energy and momentum deposition rates of supernovae over the entire galaxy volume is most likely lower than, or at most comparable to, those deposited by the wind. Moreover, the wind deposit its energy and momentum over a volume that is much smaller than the volume over which star formation takes place (at least until it expels the entire gas from the galaxy; see section \ref{sec:Msigma}) and therefore we expect it to dominate the gas dynamics in the regions relevant for accretion.  

Finally, we assumed that the wind is not highly collimated and that the ambient density is roughy spherical, as expected in high redshift bulges. If one of this assumptions is not satisfied then the wind may escape the galaxy without depositing 
its entire energy in the bulge.  In particular, in cases where the gas distribution in the galaxy is highly aspherical
(e.g., as in disk galaxies) the wind will go through a path of least resistance and may not be able to halt accretion
of sideways inflows. 

Recently, \cite{negri2017} published results of numerical simulations of accretion on galactic scales that resolve the SMBH sphere of influence. They deposit a fraction of the energy inflowing through the innermost radius resolved in the simulation back into the galaxy in the form of a conical wind with an opening angle of $45^\circ$, taking into account the effects of the processes that we ignore (multiphase structure, star formation, supernovae feedback, etc.). They find that the accretion rate is limited neither by the Bondi nor by the Eddington limits. Instead, in line with the arguments given here, it is the wind feedback that limits the accretion rate. Whenever the accretion rate is too high, the wind inflates a shocked bubble that chokes the accretion. This leads to accretion cycles that set the average 
accretion rate nearly constant. 

\section{Supercritical accretion}\label{sec:supercrit}

The above analysis implicitly assumes that all of the mass being transferred from the outer parts of the  galaxy to 
the accretion disk around the SMBH ultimately reaches the inner regions of the disk, located within some radius $r_{in}$, wherefrom roughly half of the mass is 
absorbed by the SMBH and half is ejected as winds.   It has been argued \citep[e.g.,][and references therein]{begelman2012}, 
that radiatively inefficient accretion flows (either highly sub or super Eddington), may eject  most of their mass at some large disk radius, $r_{out} \gg r_{in}$, before it reaches $r_{in}$, so that $\dot{M}_w \gg \dot{M}_{BH}$ (i.e., $\alpha \ll 1$). These arguments led to the conclusion that the rate at which a black hole can absorb mass cannot largely exceed the Eddington limit, even
if mass is being supplied to the outer disk radius $r_{out}$ at highly supercritical rates  \citep[but see e.g.,][for other considerations]{volonteri2015,begelman2017}.
If true, then as long as $\dot{M}_c \gg \dot{M}_{Edd}$ the actual accretion rate into the black hole, $\dot{M}_{BH}$, may be considerably smaller than the critical rate $\dot{M}_c$ obtained from
Eq. (\ref{acrt_rate}) upon setting $\alpha\sim1$. However, previous derivations of inflow/outflow solutions ignored potential effects of the environment, and in particular the possibility 
that the interaction of outflows ejected from outer disk radii with ambient gas on galactic scales may affect the structure of the accretion flow.  As argued below, it is quite likely that 
during the supercritical accretion phase the pressure of the ambient gas will prevent extraction of outflows from large radii, forcing the inflowing matter to reach $r_{in}$.

Suppose that mass is supplied to the disk at some radius $r_{out} \gg r_{in}$ at a rate $\dot{M}_{ac}$, and that most of it is ejected as 
winds before reaching the SMBH, so that, $\dot{M}_{BH} \ll \dot{M}_{w}\simeq \dot{M}_{ac}$.   
Now, the velocity of an outflow emanating from radius $r>r_{in}$ is a fraction $\eta \lesssim 1$ 
of the corresponding Keplerian velocity,  $v_w(r)=\eta v_K=\eta c(r_g/r)^{1/2}$, where $r_g$ is the SMBH gravitational radius.
As explained above, the wind will break out provided its momentum flux, $\dot{M}_{w}v_w\simeq \dot{M}_{ac} c\eta (r_g/r)^{1/2}$, 
exceeds $f_g\sigma^4/G$ (see Eq. (\ref{mom_balance})), or
\begin{equation}
\dot{M}_{ac} \gtrsim\eta^{-1}\left(\frac{r}{r_g}\right)^{1/2}\,\frac{f_g\, \sigma^4}{c\, G}= \left(\frac{r}{r_{in}}\right)^{1/2}\alpha^{-2}\dot{M}_c \gtrsim \dot{M}_c,
\label{dotM_w}
\end{equation}
where we adopt $v_w/c=\alpha\sqrt{2\epsilon}=\eta (r_g/r_{in})^{1/2}$ for winds expelled within $r_{in}$.  Equation (\ref{dotM_w}) implies that if $\dot{M}_{ac} \lesssim \dot{M}_c$ then 
winds ejected from radii $r>r_{in}$ do not have high enough momentum to overcome the pressure of the inflowing gas at large scales.  
The most plausible situation in our view is that the pressure of the inflowing matter prevents the formation of transonic winds from radii $r > r_{in}$, ultimately leading to accumulation of hot gas 
that applies pressure on the disk and forces the inflowing matter to reach $r_{in}$, thereby giving rise to accretion into the SMBH at a rate
 $\dot{M}_{BH} \approx \dot{M}_{ac}$  also during the super-Eddington phase. 
In this scenario the accretion is essentially regulated by the feedback of the wind ejected from $r_{in}$, so that on average $\dot{M}_{BH} \approx \dot{M}_{c}$. This scenario is not the only possible outcome. For example, the unbound gas which is ejected at $r > r_{in}$ may build up above the disk, become self gravitating and form stars. In such a case 
most of the mass that arrives at $r_{out}$ will be lost before reaching the SMBH (conceivably in the form of a breeze, as discussed in \citealt{begelman2012}), rendering
the accretion rate significantly lower than $M_c$. Note, however, that this 
would require the inflated gas to be shielded, otherwise it will be photoionized by the quasar radiation and maintained hot.  At any rate, we naively expect that the interaction of the outflow with gas on scales much larger than the disk scale will affect the disk structure and accretion rate in the super Eddington regime. Further analysis is required to assess what fraction of the inflowing gas ultimately gets absorbed by the SMBH.

If, as we expect, the accretion rate is dictated by equation (\ref{acrt_rate}) both in the sub and super Eddington phases, then the SMBH experiences roughly a constant accretion rate throughout its evolution. This means that the accretion mode switches from highly super-Eddington to sub-Eddington after approximately the 
Eddington e-folding time, $t_E\simeq 45$ Myr.  It further implies a growth time of about $150$ Myr for a $3 \times 
10^{9}~M_\odot$ SMBH, independent of its seed mass, thereby alleviating the seed size problem \citep[][ and references therin]{volonteri2010}.  Note that this conclusion holds also if during the supercritical accretion phase the rate at which the SMBH absorbs matter is well below $\dot{M}_c$, but still well in excess of the Eddington limit by virtue of 
the pressure applied on the disk through the interaction of the wind with the bulge gas.

Finally, we note that due to photon trapping, the radiative luminosity of super-Eddington 
flows (either inflows or outflows) cannot largely exceed the Eddington limit (e.g., \citealt{begelman1978}). 
This means that the observed Eddington ratio of SMBHs during the supercritical accretion phase is expected 
to be near unity, in agreement with the data in Figure \ref{fig:f1}.  At any rate, for the ambient density
profile adopted here these sources are expected to be obscured.  Note, that if a non-negligible fraction of the BH mass is accreted in the super-critical phase, as we suggest here, then not all the gravitational energy released in the accretion is released in the form of photons. This effect should be considered when one attempts to estimate the radiation released during the formation of SMBHs in the Universe along lines similar to the Soltan argument \citep{Soltan1982}.

\section{Termination of accretion phase and the $M_{BH}-\sigma$ relation}\label{sec:Msigma}

The observed SMBH masses in the local universe show strong correlations with the properties of the bulges in which they reside \citep[][ and references therein]{kormendy2013}. Past discussions of the feedback of AGN winds on the SMBH growth focused on 
its effect on the final SMBH mass. It has been argued that the SMBH will grow up to the point at which the energy \citep{silk1998} or momentum \citep{king2003} deposited by the winds becomes large enough to expel the entire gas in bulge. \cite{silk1998} find that if the shocked wind bubble does not cool efficiently and the bulge gas is driven by energy deposition then the final SMBH mass satisfies $M_{BH} \propto \sigma^5$. \cite{king2003} finds that if the shocked wind cools rapidly and the outflow is momentum driven then $M_{BH} \propto \sigma^4$.  
Both of these results are roughly compatible with the observed scalings of the $M_{BH}-\sigma$ relation \citep[e.g][]{ferrarese2000,gultekin2009,mcconnell2013}, however the normalization derived in \cite{silk1998} for energy driven outflow is vastly lower  than the observed one, while the in the case of momentum driven outflow it is consistent with the observed relation.   Rough estimates confirms that for typical parameters the shocked wind bubble is likely to cool efficiently up to a rather large radius  \citep{king2003}.   The scaling derived by both
\cite{silk1998} and \cite{king2003} is a consequence of their assumption, that the mechanical output of the SMBH is proportional to its Eddintong luminosity at all times. 
However, in the preceding sections we argued that the actual power of the wind is most likely limited by $\dot{M}_c$ and not $\dot{M}_{Edd}$, especially in late stages of the SMBH growth when  $\dot{M}_c < \dot{M}_{Edd}$.   This should yield a somewhat different scaling.
In what follows we estimate the overall growth time and final mass of the SMBH set by the AGN wind feedback we obtained above, assuming that this is the process that dominates the expulsion of gas from the bulge.

If the shocked wind bubble cools rapidly (i.e., momentum driven outflow), then the wind expels all the gas in the bulge when $M_w v_w\simeq f_g M_{bl} \sigma$, where 
$M_{bl}=10^{12.5}M_{bl,12.5}~M_\odot$ is the {\it total} (including dark matter) bulge mass. By employing Eq. (\ref{acrt_rate}) we find that the time at which this occurs is about
\begin{equation}\label{acrt_time}
t_f\simeq \frac{G M_{bl}}{\sigma^3}\simeq 300\, M_{bl,12.5}\sigma_{350}^{-3}\quad Myr,
\end{equation}
after which the accretion rate drops sharply, suppressing further black hole growth.   This yields a final mass of
\begin{equation}\label{final_mass}
M_{BH}\simeq t_f \dot{M}_c \simeq 
 5 \times 10^{9} \alpha \frac{f_g}{0.2}\left(\frac{\epsilon}{0.01}\right)^{-1/2} M_{bl,12.5} \sigma_{350}\quad M_\odot.
\end{equation}
Assuming that the total bulge mass at high redshift is proportional to the stellar mass at low redshift, $M_{bl,*}$, and using the Faber-Jackson relation, which based on recent observations yields $M_{bl,*} \propto \sigma^{3.5}$ (e.g., \citealt{gallazzi2006}), 
we obtain $M_{BH} \propto \sigma^{4.5}$, consistent with the observed M$_{BH}$-$\sigma$ relation. If, as naively expected, the stellar bulge mass is a factor of a few smaller than the total bulge mass at high redshift then both the normalization of the $M-\sigma$ relation and the Magorrian relation, $M_{BH} \sim 10^{-3} M_{bl,*}$, are obtained.  The relation obtained in Equation \ref{final_mass}, $M_BH\propto M_{bl}\sigma$, seems more fundamental in our model and can, perhaps, be tested with appropriate data.

We emphasize that rapid cooling of the wind bubble over time $t\sim t_f$ is crucial.   It can be readily shown that  had 
cooling become inefficient after time $t<<t_f$ the final SMBH mass would 
have been smaller (by roughly a factor of $\sigma/v_w$) than the value given in Eq. (\ref{final_mass}), inconsistent with observations.    
The shocked wind material cools primarily through inverse Compton scattering of radiation emitted by the central AGN and free-free emission.  
Naive estimates indicate that inverse Compton cooling is effective during the supercritical phase, but marginal during the sub-Eddington accretion phase. 
The cooling rate may be further suppressed if the cooling time is shorter than the electron-ion equilibration time, in which case a two-temperature state forms  (Faucher-Giguere \& Quataert 2012).
Whether the shocked wind indeed develops a two-temperature structure depends on the microphysics of energy exchange 
between ions and electrons, which is poorly understood at present.   On the other hand, 
beaming of the quasar radiation inside the wind funnel, mixing of shocked wind and shocked ambient matter 
at the contact due to various instabilities (as seen in Figure \ref{fig:simulation}), and conceivably other effects may enhance the cooling rate 
of the shocked wind.  Detailed analysis is needed to asses whether cooling is fast enough throughout the 
entire evolution of the system.

An interesting consequence of Eqs. (\ref{acrt_rate}) and (\ref{acrt_time}) is that while 
the accretion rate depends sensitively on the bulge properties, $t_f$ is independent 
of $f_g$ and, given the observed relation $M_{bl,*}\propto\sigma^{3.5}$, seems to depend only weakly on $\sigma$.
Thus, the picture that arises is that each SMBH accretes (on average) at a constant rate over a duration of about $300$ Myr, during which the gas fraction in its host bulge is roughly constant. After that time the emerging outflow expels most of the gas, $f_g$ 
drops significantly, and the growth of the SMBH terminates. 
If the gas in the bulge is replenished, e.g., by a subsequent galaxy merger, another episode of vigorous accretion commences. 
It will last for another several hundred Myr, during which the SMBH may double its mass.   
This process conceivably explains the decline of the maximal luminosity (i.e., highest accretion rate) of the AGN sample in Fig. 1 at redshifts $z<2$. 
We suggest that the plateau in the accretion rate of SMBHs with the highest masse
at $z >2$ indicates an ongoing activity of the most massive bulges,  e.g., galaxy mergers, that starts diminishing at $z\simeq 2$, after which all bulges gradually become devoid of gas and the average accretion rate decreases over a time scale of about $1.5$ Gyr, as seen in the middle panel of Fig. 1.    We wish to emphasize that 
the time scale over which the envelope of observed accretion rates of the entire SMBH sample declines at $z< 2$ doesn't necessarily have to be
similar to the time it takes a single SMBH to shut off its accretion, but rather may  depend on other (cosmological) factors, particularly the merger rate. 
It does have to be longer than the characteristic time of a single accretion episode predicted by our model, as indeed seems to be the case. Note that if there where SMBHs that accrete continuously at the maximal observed rate for 1.5 Gyr, we would expect the presence of SMBHs with masses significantly larger than those of the most massive SMBHs observed.

\section{Summary}
In this paper we show that regulation, by mechanical feedback, of the rate at which SMBHs accrete matter in young galaxies can account 
for the nearly constant accretion rate, of a few tens of $M_\odot$ yr$^{-1}$, and Eddington ratios observed in high redshift AGNs, as well as the $M_{BH}-\sigma$ relation 
measured in the local Universe.  Our analysis demonstrate that under reasonable assumptions the rate at which a SMBH absorbs mass 
during its cosmic evolution is limited by the interaction of winds expelled from the inner regions of the accretion flow with 
the ambient gas in the bulge. 
This limit is independent of the SMBH mass and is sensitive mostly to the bulge properties (dispersion velocity and gas fraction). 

The regulated accretion rate we derived implies a transition from highly super-Eddington accretion at early stages
to sub-Eddington accretion at times $t>t_{Edd}=45$ Myr.  We argued that the interaction of outflows expelled from 
large disk radii during the supercritical accretion phase leads to accumulation of the unbound gas 
above the disk, that in turn exerts pressure on the disk and forces the inflowing matter
to ultimately reach the inner disk regions, wherefrom the fast winds responsible for the feedback are expelled.
This gives rise to a much larger accretion rate than expected in standard inflow/outflow solutions.  We also
noted that our conclusions do not strongly depend on the exact value of the supercritical accretion rate 
as long as it is sufficiently above the Eddington limit.

Under the assumption that the shocked wind cools rapidly, we deduced that the overall evolution time of the SMBH is 
about $300$ Myr, weakly dependent on bulge properties.  After that time the wind expels the entire gas in the bulge. 
The final SMBH mass thereby obtained scales as $M_{BH}\propto M_{bl}\sigma\propto \sigma^{4.5}$, in agreement with
the observed $M_{BH}-\sigma$ relation, but in difference to previous results.  The normalization we obtained is 
further consistent with the Magorian relation, roughly $M_{BH}\sim 10^{-3}M_{bulge}$.  We note that in our model
the relation $M_{BH}\propto M_{bl}\sigma$ is more fundamental than the usual relation and, in principle, can be tested observationally.

We note that while here we consider a configuration that is most appropriate  for high redshift proto-galaxies (spherical massive gas rich bulges), regulation of the accretion rate by AGN wind feedback may be relevant also to other settings and worth being considerated also in other galaxy types. 
 
\section*{Acknowledgements}
We thank Hagai Netzer and Benny Trakhtenbrot for enlightening discussions and help,  
and Mitch Begelman for useful comments.    We are grateful to Meir Zallig-Hess for providing 
us the results of the simulations.   We also thank the anonymous  referee for a constructive 
criticism and comments that improved the presentation.
Support by The Israel Science Foundation (grant 1277/13) is acknowledged. 


\end{document}